\documentclass[twocolumn,showpacs,preprintnumbers,amsmath,amssymb]{revtex4}
\usepackage{graphicx}
\usepackage{dcolumn}
\DeclareMathAlphabet{\mathpzc}{OT1}{pzc}{m}{it}
\newcommand{\gtsim}{\mbox{{\raisebox{-0.4ex}{$\stackrel{>}{{\scriptstyle\sim
}}
$}}}}
\newcommand{\ltsim}{\mbox{{\raisebox{-0.4ex}{$\stackrel{<}{{\scriptstyle\sim
}}
$}}}}
\begin{document}

\title{Separation of energy scales in the 
kagome antiferromagnet TmAgGe: 
a magnetic-field-orientation study up to 55 T }

\author{P. A. Goddard$^{1,2}$, J. Singleton$^2$, 
A. L. Lima-Sharma$^2$\footnote{Now at
RIKEN, 2-1 Horosawa, Wako-shi, 351-0198, Japan}, 
E. Morosan$^3$\footnote{Now at
Department of Chemistry, Princeton University, NJ~08540}, S. J. Blundell$^1$, 
S. L. Bud'ko$^3$ and P. C. Canfield$^3$.}

\affiliation{$^1$Clarendon Laboratory, 
Oxford University, Parks Road, Oxford, OX1 3PU, UK}
\affiliation{$^2$National High Magnetic Field Laboratory, 
Los Alamos National Laboratory, MS-E536, 
Los Alamos, NM~87545}
\affiliation{$^3$Ames Laboratory and Department of Physics and Astronomy, 
Iowa State University, Ames, IA 50011.}

\begin{abstract}
TmAgGe is an antiferromagnet in which the spins
are confined to distorted kagome-like planes at low temperatures. 
We report angle-dependent measurements of the magnetization $M$
in fields of up to 55~T
that show that there are two distinct and separate energy
scales present in TmAgGe, each responsible for
a set of step-like metamagnetic transitions; 
weak exchange interactions and 
strong crystalline electric field (CEF) interactions. Simulations of
$M$ using a three-dimensional, free-energy minimization
technique allow us to specify for the first time the physical
origin of the metamagnetic transitions in
low, in-plane fields. We also show that
the transitions observed with the field perpendicular
to the kagome planes are associated with the CEF-split multiplet of Tm.
\end{abstract}


\maketitle

\begin{figure}[t]
\includegraphics[width=5cm]{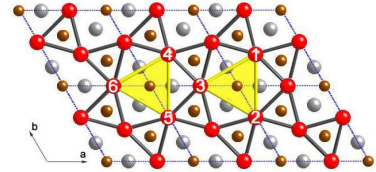}
\vspace{-4mm}
\caption{Projection of the distorted kagome structure of TmAgGe in the
$ab$-plane. Tm: red, Ag: silver, Ge: gold. The model in this paper
has spins 1--6 as its
repeating structure. 
Nearest neighbors 1, 2 $\&$ 3 and 4, 5 $\&$ 6 have
AF interactions; 3 has FM interactions with
4 and 5; because every second triangle of spins is
considered equivalent, 2 has FM interactions with 4 and
6. No FM interactions act on 1 in this model.
\label{fig1}}
\vspace{-4mm}
\end{figure}

Interest in antiferromagnetic (AF) systems based on
structures with possible frustration, such
as triangular, kagome or pyrocholore lattices, has
recently burgeoned,
as such materials often exhibit novel
cooperative phases~\cite{moessnerramirez}. 
One example, exhibiting complex
metamagnetic behavior, is TmAgGe,
a member of the
{\it R}AgGe ({\it R}=Tb, Dy,
Ho, Er, Tm and Y) family of metallic
compounds. TmAgGe adopts a layered, 
distorted-kagome lattice
similar to the ZnNiAl structure (Fig.~\ref{fig1})~\cite{baran,morprb}. 
In this paper, we show that a
model based on
a six-spin repeating structure
can account for the majority
of magnetic data on TmAgGe, including the
field ($H$) positions of the metamagnetic
transitions and the magnitude of the magnetization $M$,
when $H$ is applied within the kagome planes.
Though TmAgGe exhibits few of the features
conventionally associated with frustration,
the model shows that it conforms to the
fundamental definition of a frustrated 
system~\cite{moessnerramirez}:
the geometry
of the lattice precludes the simultaneous minimization
of all of the interaction energies, in this case
antiferromagnetic (AF) next-nearest-neighbor (NNN)
and ferromagnetic (FM) nearest neighbour (NN) exchange
with a characteristic energy scale $\sim 4$~K.
We also report the first observation of a series of
high-field metamagnetic transitions when $H$
is approximately parallel to $c$;
these are due to field-induced level crossing within
the crystalline-electric-field (CEF) split Tm$^{3+}$
$J=6$ (4f$^{12}$, $^3$H$_6$) multiplet. 
Here, the energy scale
is $\sim 100$~K. 
Based on these two energy scales (i.e. exchange $\sim 4$~K,
CEF $\sim 100$~K), we can produce a complete
quantitative phase diagram for TmAgGe.

\begin{figure}[t]
\includegraphics[width=7.0cm]{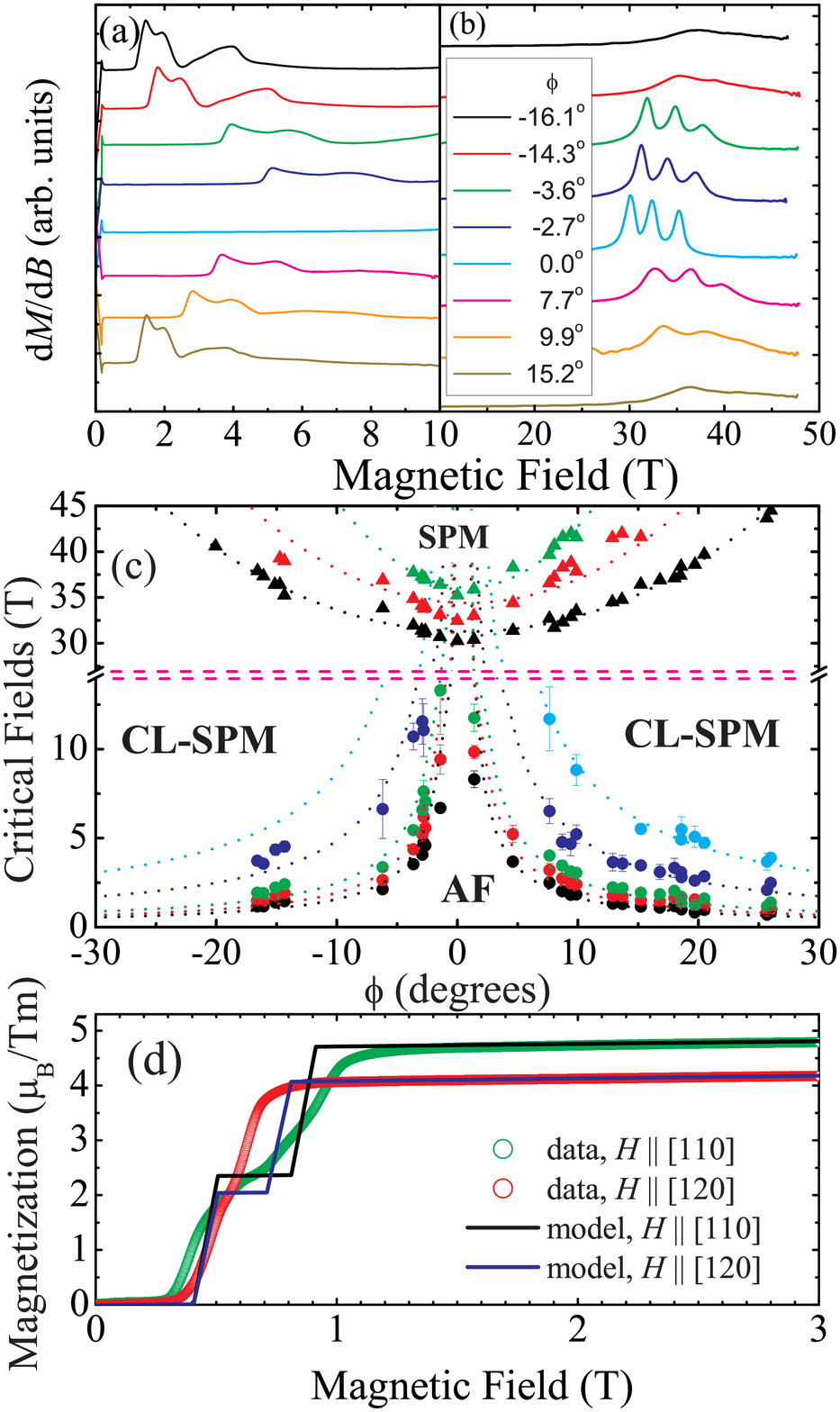}
\vspace{-5mm}
\caption{(a),(b)~Magnetic suceptibility 
$\chi={\rm d}M/{\rm d}H$ of TmAgGe
($T \approx 500$~mK) measured in pulsed magnetic fields 
at various values of the
inclination angle $\phi$. 
Peaks in $\chi$ correspond to rapid rises of 
$M$ (transitions) between broadened
steps. Data are offset for
clarity. (c)~Critical fields of the magnetic transitions
($T \approx 500$~mK) plotted as a function of $\phi$.  
AF = antiferromagnet; CL-SPM = crystal-field limited
saturated paramagnet; SPM = saturated paramagnet. 
Dotted lines are
best fits to $1/|\sin \phi|$ (low fields) 
and $1/|\cos \phi|$ (high fields).
(d)~$M$ predicted by the model (see text)
with $H$ along the
[110] and [120] directions
compared with experimental pulsed-field data.
} 
\label{fig2}
\vspace{-5mm}
\end{figure}

Much of the essential physics is
captured in Fig.~\ref{fig2}.
Fig.~\ref{fig2}(a) shows the metamagnetic
transitions (peaks in $\chi$, corresponding
to steep rises in $M$) at
several values of $\phi$, the angle between $H$ and the
$c$-axis. These low-$H$
transitions, determined by the $\sim 4$~K
energy scale, depend only on the component
of $H$ parallel to the $ab$-planes. 
At higher $H$, the new series of
transitions is observed (Fig.~\ref{fig2}(b)); 
as shown below in the discussion of
Fig.~\ref{fig6}, these result in $M$
approaching its full saturated value of $7\mu_{\rm B}$ per Tm$^{3+}$
ion. These transitions,
determined by the $\sim 100$~K energy scale, 
depend only on the
component of $H$ perpendicular to the $ab$-planes. 
The $(H, \phi)$ phase diagram is
in Fig.~\ref{fig2}(c); note that the low- and high-$H$
transitions can be seen in the same
field sweep, showing that they are distinct
phenomena, separated at all $\phi$.

Oriented single
crystals of TmAgGe (for growth details, see~\cite{mor04}), 
$\sim 0.7\times0.7\times2$~mm$^3$,
were used 
in compensated-coil
$M$ measurements in a 65~T pulsed magnet
at NHMFL.
In one probe, the sample can be 
inserted into and extracted from the coil
{\it in situ}, enabling a measurement of
the sample's $M$. 
In the second probe, the sample is
in a coil that
tilts so that the angle-dependence of
$\chi ={\rm d}M/{\rm d}H$ can be
recorded. This was used to obtain the data 
in Fig.~\ref{fig2}. 
The inclination is deduced by 
comparing the voltage induced in an
empty coil on the tilting platform, and that from another
empty, static coil. 
Both probes were placed in $^3$He
cryostats ($T~\gtsim~500$~mK).
Pulsed-field data were calibrated 
against results from
Quantum Design MPMS systems ($\mu_0 H \leq 7$~T;
$T \geq 2$~K).

The sharp features in $\chi$ observed in Fig.~\ref{fig2}(a)
correspond to steep increases of $M$ on either
side of rounded plateaux.
Examples are shown
in Fig.~\ref{fig2}(d) for $\phi=90^{\circ}$ and
$\theta =0$ ($H|| [110]$) and $\theta =30^{\circ}$
($H || [120]$); here, $\theta$ is
an azimuthal angle coordinate,
for $H$ rotating in the $ab$ plane.
In the former data,
a broadened plateau 
is centred on $M\approx 2.3\mu_{\rm B}{\rm Tm^{-1}}$
and in the latter, a smaller, rounded shelf can be discerned at
$M\approx 2.0\mu_{\rm B}{\rm Tm^{-1}}$.
These data, and others recorded at fixed $\theta$,
reproduce quantitatively the $M(H,\theta)$ measurements
in Fig.~24 of Ref.~\cite{morprb}.  
To model such $M$ data, and to 
predict the $(H,\theta,\phi)$ phase
diagram of TmAgGe, we use the following Hamiltonian:
\begin{eqnarray} 
\mathpzc{\hat{H}} &=& J_{\rm AF}\sum_{ij} ^{NNN}{\bf S}_i \cdot {\bf S}_j + J_{\rm FM}\sum_{ij}^{NN} {\bf S}_i \cdot {\bf S}_j \nonumber \\
&& + \Delta \sum_{i}(S_i^\Delta)^2 - g \mu_{\rm B}\sum_{i}{\bf S}_i
\cdot {\bf B} .
\label{hamil}
\end {eqnarray}
Here, $J_{\rm AF}>0$ and
$J_{\rm FM}<0$ are NNN AF
and NN FM exchange constants,
$S_i^\Delta$ is the component
of the $i$th spin along the local easy axis~\cite{morprb}
and $\Delta<0$ is the anisotropy energy.
The 6-spin repeating structure (Fig.~\ref{fig1}) 
is sufficient
to reproduce the majority of the features of $M(H,\theta)$.

\begin{figure}[t]
\includegraphics[width=7.5cm]{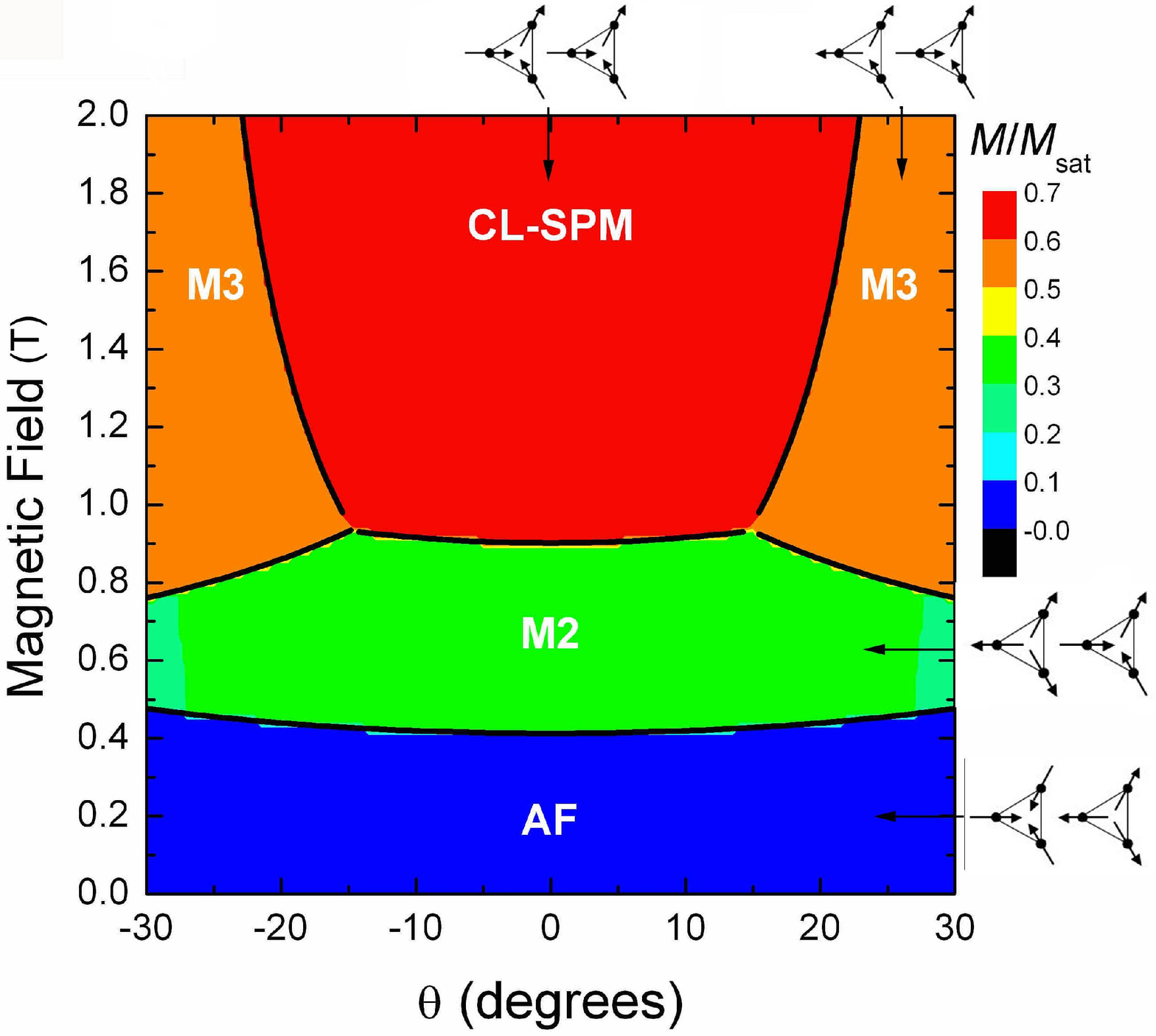}
\includegraphics[width=7.0cm]{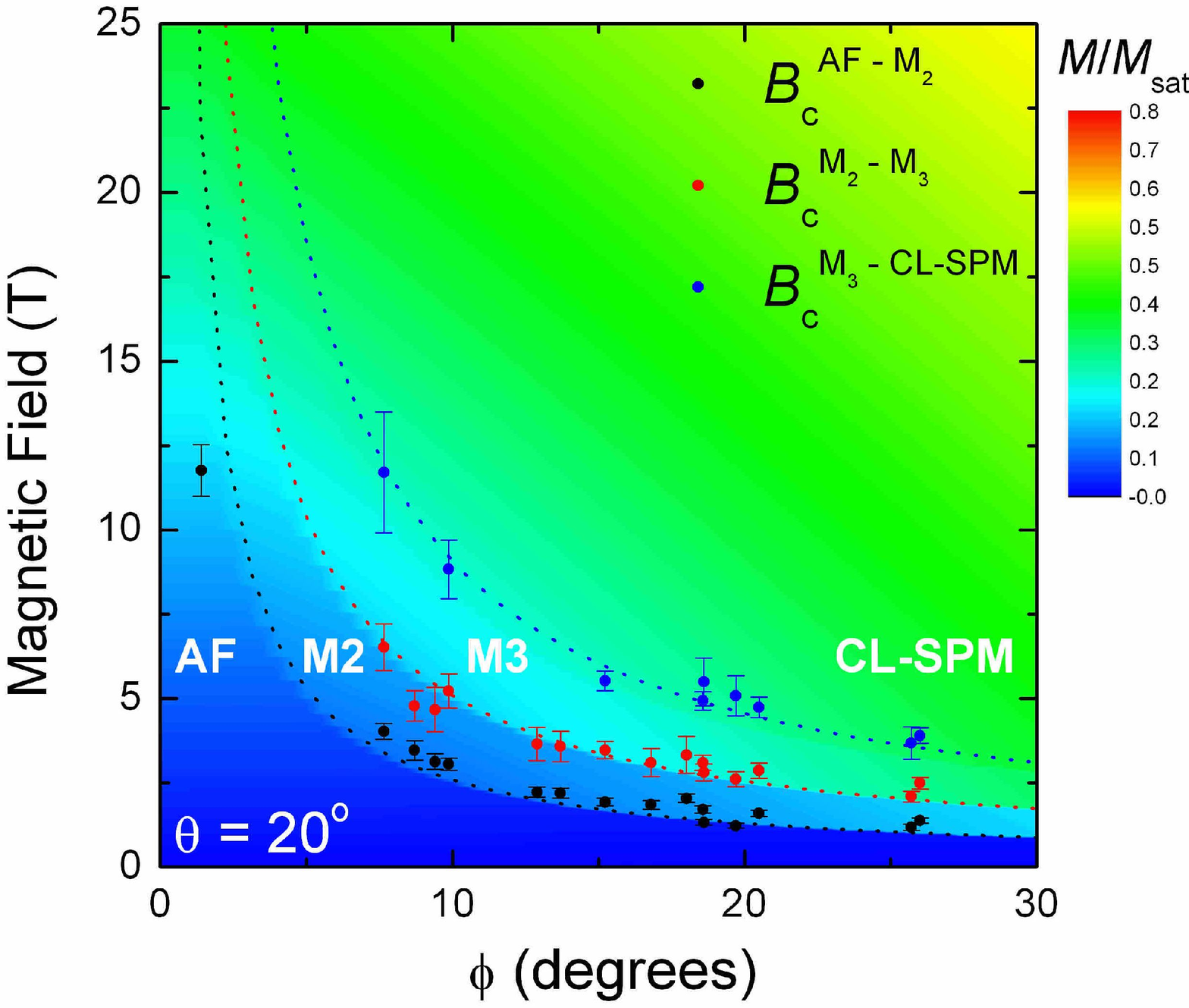}
\vspace{-5mm}
\caption{$(H,\theta)$ (a) 
and $(H,\phi)$ (b)
phase diagrams of TmAgGe
predicted by the model (see text). 
The color keys show $|M|$. 
(a)~Spin configurations are shown for each
state; 
solid lines show critical-field $\theta$-dependences
from Eq.~\ref{critter}.
The diagram agrees well with 
data in this work (e.g. Fig.~\ref{fig2}(d))
and in Fig.~25 of~\cite{morprb}.
Note that the diagram has a $\theta$ periodicity
of $60^{\circ}$.
(b)~Points are
experimental critical fields 
(Figs.~\ref{fig2}(a), (c));
dotted lines are fits of data to 
$1/|\sin \phi|$ ($\theta=20^{\circ}$).} 
\label{fig3}
\vspace{-5mm}
\end{figure}

\begin{figure}[t]
\vspace{-0.6cm}
\includegraphics[width=6.8cm]{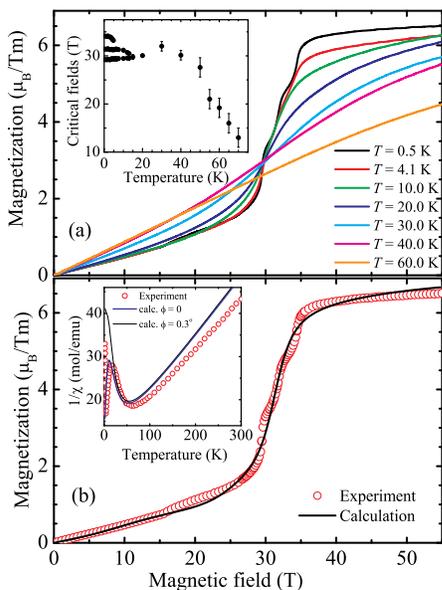} 
\vspace{-0.8cm}
\caption{(a)~$M(H,T)$ of TmAgGe ($H \parallel c$) 
showing the transitions
attributed to CEF-split energy-level crossing. 
Inset:
critical fields versus $T$.
(b)~Calculated $M$ with $H \parallel c$ found by
diagonalizing $\hat{H}_1$ (see text), and
experimental data ($T=500$~mK for both). 
The Stevens parameters were
$B_2^0=1300$~mK; $B_4^0=-3.1$~mK; $B_4^2=19$~mK; and
$B_6^0=0.0068$~mK, all the rest being zero. 
Inset:
calculated $1/\chi$ using the same $B_n^m$ at two
different directions of $H$, compared with
data obtained at 1~T~\cite{mor04}.} 
\label{fig6}
\vspace{-5mm}
\end{figure}

Having established the interactions and the number of spins
involved, the simulation involves using the 
{\it downhill simplex method}~\cite{neld65,numrecip} 
to find the
minimum free-energy spin configuration
for a particular value and direction of $H$.
The parameters $J_{\rm AF}$, $J_{\rm FM}$
and $\Delta$ are then adjusted
to quantitatively match
$M(H,\theta,\phi)$ data ({\it e.g.} Figs.~\ref{fig2}(d)
and \ref{fig3}(b)), a process that
provides a tight constraint of the values.
Note that in contrast to previous phenomenological
geometrical models of TmAgGe~\cite{morprb}, 
there is no need
to impose a starting spin configuration;
our model automatically finds a reasonable arrangement.

Simulations are shown with data in Fig.~\ref{fig2}(d),
and corresponding spin configurations
are given in the $(H,\theta)$
phase diagram ($\phi = 90^{\circ}$)
of Fig.~\ref{fig3}(a) 
for the various states.
In the AF groundstate, the spins on each 
triangular plaquette lie in the
non-colinear directions dictated by the 
anisotropy, with spins
pointing inwards and outwards 
alternately on adjacent triangles in
order to minimize the energy of the FM term in the Hamiltonian. 
As $H$ increases along [110] ($\theta=0$) the
spins remain parallel to their easy axes, 
but three of the spins flip
towards the direction of $H$, while two flip back to minimize the
AF term's energy.
This forms the M$_2$ state~\cite{morprb}, corresponding
to the plateau in $M$ in the model and 
the broadened step in the data. As $H$
increases further, the AF interactions are overcome 
and the spins rotate
towards $H$, leading to the crystal-field-limited
saturated paramagnetic (CL-SPM) state.
For $H || [120]$ ($\theta=30^\circ$),
the M$_2$ state again forms (plateau in model,
inflexion in data) but the CL-SPM state is
not realized, as one of the spins is perpendicular to, 
and so cannot
couple with, $H$. Instead, the FM interactions force the
moments into the M$_3$ state.

In addition to the absolute size of $M$ within the
various states (color scale), 
Fig.~\ref{fig3}(a) shows the
$\theta$-dependences of the critical fields
marking the boundaries between the spin 
configurations (thick lines).
These are obtained by
equating the free energies of the states, yielding: 
\begin{eqnarray} 
B_{\rm c}^{\rm AF \rightarrow M_2}&=& \beta J_{\rm AF}/ \cos \theta;\nonumber \\
B_{\rm c}^{\rm M_2 \rightarrow M_3} &=&\beta(2J_{\rm AF}-J_{\rm FM}) /(2\cos( \theta-\pi/3)); \nonumber \\ 
B_{\rm c}^{\rm M_2 \rightarrow CL-SPM}&=&\beta (J_{\rm AF}-J_{\rm FM}) /\cos \theta; \nonumber \\  
B_{\rm c}^{\rm M_3 \rightarrow CL-SPM} &=& -\beta J_{\rm FM} /(2\cos (\theta+\pi/3)).
\label{critter}
\end{eqnarray}
Here, $\beta=J/g_J\mu_{\rm B}$,
where $J=6$ and $g_J=7/6$ are respectively the total angular momentum 
quantum number and Land\'e g-factor
for Tm$^{3+}$. 

Eqs.~\ref{critter} are similar
to those extracted from $M(H,\theta)$ data in Ref.~\cite{morprb},
but with explicit prefactors;
comparison of Eqs.~\ref{critter}
with the experimental critical fields observed in the
current study and in Ref.~\cite{morprb} 
forms a tight constraint on the model parameters $J_{\rm AF}$
and $J_{\rm FM}$. 
We find $J_{\rm AF} = 0.054 \pm 0.003$~K and 
$J_{\rm FM} = -0.064 \pm 0.003$~K. The energy
ranges of the first two terms 
in Eq.~(\ref{hamil})
are $\pm J^2J_{\rm AF}$ and $\pm J^2J_{\rm FM}$; thus the energy
scale at which the exchange interactions are important is $\sim 4$~K.
Not only does this energy scale account for the sizes
of the critical fields of the 
low-field metamagnetic transitions (Figs.~\ref{fig2}(a),(d));
it is also responsible for  
the $H=0$ transition at $T_{\rm M}\approx 4.2$~K
into the state that we have labelled.
Although the AF state has zero moment, the model shows that
the FM interactions are ultimately responsible for
its spin configuration.
This, and the plethora of other states
observed in the data (Figs.~\ref{fig3}(a), (b))
result from competition
of FM and AF interactions; the geometry
of the lattice precludes the simultaneous minimization
of all of the interaction energies.
In this respect, TmAgGe conforms to the most fundamental
definition of a frustrated system~\cite{moessnerramirez}. 

Whilst our 6-spin system (Fig.~\ref{fig1})
and Hamiltonian (Eq.~\ref{hamil})
describe most of the low-field
$M(H,\theta,\phi)$ data presented in this paper and 
in Ref.~\cite{morprb},
it is necessary to make two comments.
First, the rise in $M$ observed
between the low-field AFM state and the 
broadened step associated with the M$_2$
spin configuration (Fig.~\ref{fig2}(d))
is attributed elsewhere~\cite{morprb}
to a state labelled M$_1$; our model
does not predict M$_1$, instead
showing a direct AFM-M$_2$ transition (fig.~\ref{fig3}(a)).
Second, to achieve a quantitative reproduction
of all the other states observed, it is
necessary to suppress FM interactions between spin 1
in Fig.~\ref{fig1} and its NNs: if these interactions
are switched on in the simulations then the angular region over which
the M$_3$ state is observed is severely reduced and the M$_2$ state
disappears completely, in contradiction to experiments. 
The most likely explanation is that our 6-spin model
is a subset of the actual repeating magnetic structure
in TmAgGe. Support for this view comes from
the structures inferred from analysis of
$M(H,\theta)$ data in Ref.~\cite{morprb}, 
which suggest that
$\gtsim~18$ spins are required to reproduce $M$ within
the M$_1$ state.
Nevertheless, we emphasise that the 6-spin
repeating system (Fig~\ref{fig1}, Eq.~\ref{hamil})
is able to describe almost all of the data,
especially those at high $H$, and represents a 
tractable model for exploring
spin physics in layered kagome systems of this kind.

We now turn to the other
energy scale in TmAgGe, associated with
the CEF interactions. Though the moments are strongly inclined to
lie along the easy axes, the finite $\Delta$ 
means that they will
cant towards $H$, leading to the gradual
increase in $M$ as $H$ rises
seen in the CL-SPM state (Fig.~\ref{fig2}(d)).
Using this gradient, we find that $\Delta=-4.6 \pm 0.1$~K. 
This explains the large negative
Curie temperature for $H||c$ ($\Theta_{c}$);
the Zeeman and anisotropy terms dominate in Eq.~\ref{hamil}
for $H||c$, and in this limit one can show that the
high-$T$ $1/\chi$ is linear with a $T$-axis
intercept of 
$\Theta_{c}\approx -\Delta(2J+3)(2J-1)/15\approx -50$~K, in
reasonable agreement with experiment 
($\Theta_{c}\approx-76$~K~\cite{mor04}).

The third term of Eq.~\ref{hamil} has a magnitude
$\sim |J^2 \Delta|\sim 170$~K, forcing a non-colinear
or ``compromise'' structure on the Tm moments
at $T$s far in excess of those at which the AF interactions
become important. 
This compromise structure, in which the moments
lie in the $ab$ planes $120^{\circ}$ apart, 
could potentially possess 
a number of different
degenerate spin configurations~\cite{moessnerramirez}. 
However, in TmAgGe the model shows that
the degeneracy is lifted
by the FM NN interactions, leading to the 
states at low $T$s and the magnitude and sign of 
$\Theta_{ab}\approx +7.5$~K~\cite{mor04}. 
As mentioned above,
these states are then accessible using small
in-plane ($\phi=90^{\circ}$) $H$s, 
accounting for the metamagnetic
transitions (Figs.~\ref{fig2}(d) and \ref{fig3})~\cite{footnote}. 

Fig.~\ref{fig3}(b) shows the $(H,\phi)$
phase diagram of TmAgGe simulated using
the above parameters, together with experimental data. 
In agreement with experiment, the
relatively large $|\Delta|$ confines
the spins to the $ab$ planes, so that the
low-$H$, metamagnetic transitions depend only on the component of
$H$ in the planes; the critical fields
scale as $1/|\sin \phi|$~\cite{xxx}. 
Note that the simulation uses $\theta=20^{\circ}$,
in agreement with the nominal orientation of the sample.

We now turn to the high-$H$ metamagnetic transitions
observed when $H$ is applied out of the $ab$ planes
(Fig.~\ref{fig2}(b)). As mentioned before,
peaks in $\chi$ correspond to steep rises
in $M$; this is shown in Fig.~\ref{fig6}(a) for $H || c (\theta=0)$.
Eq.~(\ref{hamil}) predicts that for
$H \parallel c$, the
low-$T$ $M$ should increase linearly in $H$ with a
gradient given by $g_J\mu_{\rm B}/2\Delta$, reaching its saturated
value of 7$\mu_{\rm B}$ per Tm$^{3+}$ ion at $\mu_0H \approx 70$~T.
The experimental $M$ in Fig.~\ref{fig6}(a) does
increase linearly at low $H$ (albeit at low $T$
with a smaller gradient than predicted); however, 
at $\mu_0H \approx 35$~T
the steep rise in $M$ mentioned above is observed, and
$M$ approaches its saturated value.
This is similar to what has been
observed in high-$H$ measurements of HoNi$_2$B$_2$C~\cite{abliz}. 
For $T~\ltsim ~20$~K, there is a two- or 
three-fold structure within
the rise in $M$; at higher $T$, these structures
are not resolved, but a single transition in $M$ is visible
to $T\approx 70$~K (Fig.~\ref{fig6}(a), inset).
$T$s of 20~K and 70~K are several times higher
than $T_{\rm M}\approx 4.2$~K, 
indicating that the transition(s) in $M$
in Fig.~\ref{fig6} cannot easily be attributed to
exchange interactions. 
Indeed, as one tilts $H$ away from $c$ (Figs.~\ref{fig2}(a), (b)),
the low-$H$ metamagnetic transitions
that can be confidently attributed to the competing
FM and AF exchange interactions are observed
within the same field sweeps as the transition(s)
at $\mu_0H\approx 35$~T; this strongly
suggests that the mechanisms for the
low- and high-$H$ transition(s) are
distinct.

The field of $\mu_0H\approx 35$~T
suggests an energy scale for the high-$H$
transitions similar to
the anisotropy term in Eq.~\ref{hamil}. 
The likely candidate is CEF splitting of the $J = 6$
multiplet, which causes the 
easy-axis anisotropy seen at low $T$. 
As stated above,
such splittings would be of the order of 
$|J^2\Delta| \sim 170$~K (Eq.~\ref{hamil}, third term).
When the Zeeman
energy $=g_J\mu_{\rm B}JB$ becomes of this order, 
level crossings are expected; this occurs when $\mu_0H \approx 36$~T,
in agreement with the data.
The Tm ions are located at sites with orthorhombic 
symmetry $C_{2v}~(2mm)$~\cite{baran} and so
we consider the Hamiltonian
$\hat{H}_1=
\sum_{n=2,4,6}~\sum_{m=0}^n B_n^m O_n^m
-g\mu_{\rm B}\sum_i{\bf S}_i\cdot{\bf B}$,
a sum of CEF~\cite{walter84} and Zeeman terms,
where the $B_n^m$ and $O_n^m$ are the CEF parameters and Stevens
operators, respectively. The $O_n^m$ with $m\neq0$ contain the
angular momentum raising and lowering operators; thus, if the
corresponding CEF parameters are large enough, 
these terms will lead to
substantial mixing of the $J_z$ energy levels.

By diagonalizing $\hat{H}_1$ and using expressions
for $\chi$ and $M$ from Ref.~\cite{abliz}, we
find values for the Stevens parameters such that the low-$T$
$M$ with $H\parallel c$
is a reasonable match to the experimental data
(Fig.~\ref{fig6}(b)). The low-$H$ $c$-axis $M$ is well
reproduced, as is the transition to saturation around 35~T.
However the three steps within this transition are not found
in the calculations. The inset in Fig.~\ref{fig6}(b) shows the
calculated low-field $1/\chi$
versus $T$. 
Whilst the calculations reproduce
the inset data qualitatively, the quantitative agreement
is less good than in Fig.~\ref{fig6}(b), partly because
at low $T$ the calculated $1/\chi$ 
is very sensitive to $\phi$;
the same is true to a lesser extent
of the experimental data. 
Note that
the $B^m_n$ values in the caption to
Fig.~\ref{fig6} merely
represent a possible solution to the problem;
our intent here is
to show that CEF splitting of the $J = 6$ multiplet {\it alone} can
account for the primary features of the $T$
dependence of the $c$-axis $\chi$ and $M(H)$.
A further constraint on the $B_n^m$ parameters
must await inelastic neutron scattering experiments.

This work is supported by the U.S. Department of Energy (DoE), 
NSF, the State of Florida, 
EPSRC (UK) 
and the Oxford Glasstone Fellowship Scheme. 
Ames Laboratory is operated for DoE
by Iowa State University under Contract W-7405-ENG-82.
We thank
Susan Cox and Neil Harrison for helpful suggestions.


\begin{thebibliography}{99}
\bibitem{footnote}
An analogous result
was found for jarosites~\cite{ballou}.
Jarosites are
insulators, and so the
exchange mechanism differs from
metallic TmAgGe. Nevertheless,
the physics of competing NN and
NNN interactions is the same.
\bibitem{xxx}
Similar angle dependences have been observed for other
highly anisotropic, local-moment systems when $H$ is rotated away from
the easy axis (or plane)~\cite{budko}.
\end{thebibliography}
\end{document}